\documentstyle[twocolumn,prb,aps,epsf]{revtex}

\def\be{\begin{equation}}
\def\ee{\end{equation}}
\def\bea{\begin{eqnarray}}
\def\eea{\end{eqnarray}}

\begin{document}

\twocolumn[\hsize\textwidth\columnwidth\hsize\csname
@twocolumnfalse\endcsname

\title{Effective three-band model for double perovskites}
\author{P. Petrone and A. A. Aligia}
\address{Comisi\'on Nacional de Energ{\'\i}a At\'omica,
Centro At\'{o}mico Bariloche and Instituto Balseiro, 8400 S.C. de Bariloche,
Argentina.}
\date{Received \today }
\maketitle

\begin{abstract}
We start from a six-band model describing the transition-metal $t_{2g}$
orbitals of half-metallic double perovskite systems, such as Sr$_{2}$FeMoO$_{6}$, in which only one of the transition metal ions (Fe) contains
important intratomic repulsion $U_{Fe}$. By eliminating the Mo orbitals
using a low-energy reduction similar to that used in the cuprates, we
construct a Hamiltonian  which contains only effective $t_{2g}$ Fe orbitals.
This allows to treat exactly $U_{Fe}$ and most of the Fe-Mo hopping. As an
application, we treat the effective Hamiltonian in the slave-boson
mean-field approximation and calculate the position of the metal-insulator
transition and other quantities as a function of pressure or on-site energy
difference.
\end{abstract}


] 
\narrowtext

\section{Introduction}

In the simple perovskites of formula unit $AM$O$_{3}$, the transition metal
atoms $M$ form a simple cubic lattice. In the double perovskites $%
A_{2}MM^{\prime }$O$_{6}$, this lattice is subdivided into two
interpenetrating f.c.c. sublattices, in such a way that the nearest
transition metal $M^{\prime }$ lies in the sublattice occupied by $M$ and
vice versa. $A$ is an alkaline earth. Recently, the interest on these
systems has considerably increased after the finding of low-field colossal
magnetoresistance (CMR) in Sr$_{2}$FeMoO$_{6}$, which remains significant at
room temperature.\cite{koba} The system is a half metallic ferromagnet with
a Curie temperature near 450K. The mechanism leading to CMR at low fields is
believed to be intergrain carrier scattering between regions of different
orientation of magnetization, which is reduced under an applied magnetic
field. Other double perovskites systems which were studied include Ba$_{2}$%
FeMoO$_{6}$, \cite{ba} Sr$_{2}$FeReO$_{6}$,\cite{re} Sr$_{2}$FeWO$_{6}$,\cite
{w} and the alloy Sr$_{2}$FeMo$_{1-x}$W$_{x}$O$_{6}$.\cite{alloy} Sr$_{2}$%
FeWO$_{6}$ is an insulating antiferromagnetic system, and the research on
this alloy monitors the transition from metallic itinerant ferromagnetism in
Sr$_{2}$FeMoO$_{6}$ to localized-electron antiferromagnetism in Sr$_{2}$FeWO$%
_{6}$. Also, substitution of Fe by Co or Mn renders Sr$_{2}$FeMoO$_{6}$
antiferromagnetic and insulating.\cite{itoh} Ordinary ab initio calculations
fail to explain the electronic structure of Sr$_{2}$FeWO$_{6}$, obtaining a
metallic ferromagnetic ground state.\cite{fang} This result changes if the
phenomenological method called LDA+U is used.\cite{fang}

An issue related with the electronic structure of Sr$_{2}$FeMoO$_{6}$ which
is not settled yet is the valence of the transition metal ions. This is in
principle related with the metallic or insulating character, since in an
ionic picture with oxidation states O$^{-2}$, Sr$^{+2}$, Fe$^{+2}$ and a
closed shell Mo$^{+6}$, one might expect that the strong Coulomb repulsion
at Fe sites $U_{Fe}$ brings the system close to a Mott transition.\cite{ali}
While the Mo-3d chemical shift observed in optical experiments is
practically identical to that of MoO$_{3}$ (indicating formal valence Mo$%
^{+6}$), the formal valence of Fe seems to be +3.\cite{more} This apparently
contradictory result is interpreted by the authors as an indication of
covalency. M\"{o}ssbauer experiments are interpreted also as indicating
covalency.\cite{lind} Nevertheless, as we shall see, even assuming that all
O ions are O$^{-2}$, there is a certain degree of covalency in the
insulating state. Results from neutron diffraction experiments obtain
magnetic moments $\mu _{Fe}=4.1\pm 0.1\mu _{B}$ and $\mu _{Mo}=0\pm 0.1\mu
_{B}$.\cite{gar} These values are consistent with an insulating state,
suggesting that Sr$_{2}$FeMoO$_{6}$ is near a metal-insulator transition.

The metal-insulator transition has been studied theoretically applying the
slave-boson mean-field approximation (SBMFA) to a six-band model containing
the relevant Mo and Fe $t_{2g}$ orbitals, neglecting the hopping between Mo
ions and taking a simplified density of states \cite{ali}. The SBMFA permits
to treat $U_{Fe}\simeq 7eV$ in a way equivalent to the Gutzwiller
approximation.\cite{kot}

The aim of the present work is to derive an effective 3-band Hamiltonian,
eliminating the Mo sites by a suitable low-energy reduction procedure. This
procedure has been successfully used in the cuprates. After the original
proposal of Zhang and Rice \cite{zr} that in spite of Cu-O covalency, the
low energy physics of the superconducting cuprates can be described by a
one-band model, several systematic studies have derived the different terms
of this model and used it successfully to calculate several properties. \cite
{jef,sch,sb,bel,tri,fei,ros,opt} The advantages of the effective model is
that it has a smaller Hilbert space in numerical diagonalizations of finite
systems and that the largest interaction in the problem $U_{Fe}$ (or the
on-site repulsion at Cu in the case of cuprates) is treated exactly inside
an effective cell. As a consequence, one expects that approximate treatments
give better results when applied to the effective Hamiltonian rather than
the original one. This is the case of the SBMFA applied to calculate the
metal-insulator transition in the cuprates: the results are significantly
improved when applied to the effective one-band model.\cite{sb}

Our resulting low-energy Hamiltonian contains only effective Fe sites with a
reduced effective Coulomb repulsion ($U<2.8$ eV for Sr$_{2}$FeMoO$_{6}$). It
can be described as a one-band model containing three pseudospin or
``color'' components describing the $xy$, $yz$, and $zx$ $t_{2g}$ orbitals
of the 3d or 4d shell. For the simpler two-color version of this effective
model in infinite dimensions, the critical value of $U$ at the Mott
transition lies very near the exact result.\cite{roze,geor} We apply the
SBMFA to the resulting effective Hamiltonian, to study the metallic or
insulating character of the system as a function of the parameters of the
starting 6-band Hamiltonian. We also analyze the effect of pressure using 
{\it ab initio} calculations. The paper is organized as follows. In Section
II, we describe the original 6-band model and the way we obtain its
parameters using {\it ab initio} data. The effective 3-band model is derived
in Section III. Section IV contains results for the metal-insulator
transition using SBMFA. Section V is a summary and discussion.

\section{The starting Hamiltonian}

We start from a six-band model which describes the $t_{2g}$ orbitals of
minority spin of the two different transition-metal atoms in a NaCl
structure. One of them, denoted by Fe, contains an important on-site Coulomb
repulsion $U_{Fe}$, while valence electrons of the other transition metal
(like the 4d orbitals of Mo) are more extended, and as a consequence, the
on-site Coulomb repulsion can be neglected. The Hamiltonian is:

\begin{eqnarray}
&&H=\varepsilon _{Fe}\sum_{i\sigma }f_{i\sigma }^{\dagger }f_{i\sigma
}+\varepsilon _{Mo}\sum_{j\sigma }m_{j\sigma }^{\dagger }m_{j\sigma } 
\nonumber \\
&&+U_{Fe}\sum_{i\sigma <\sigma ^{\prime }}f_{i,\sigma }^{\dagger
}f_{i,\sigma }f_{i,\sigma ^{\prime }}^{\dagger }f_{i,\sigma ^{\prime
}}-[\sum_{i\sigma \delta _{\sigma }}[t_{FM}f_{i,\sigma }^{\dagger
}m_{i+\delta _{\sigma },\sigma }  \nonumber \\
&&+\sum_{j\sigma \sigma ^{\prime }\gamma }t_{MM}(\sigma ,\sigma ^{\prime
},\gamma )m_{j,\sigma }^{\dagger }m_{j+\gamma ,\sigma ^{\prime }}+\text{H.c}%
.]  \label{hs}
\end{eqnarray}
$f_{i\sigma }^{\dagger }$ creates an electron at the $d_{\sigma }$ ($\sigma
=xy$, $yz$, or $zx$) orbital at the Fe site $i$ with minority spin. $%
m_{j\sigma }^{\dagger }$ has a similar meaning for Mo site $j$. The four
two-dimensional vectors $\delta _{\sigma }$ connect an Fe site with its
nearest-neighbor (NN) Mo sites lying in the same plane as the orbital $%
\sigma $. Similarly, $\gamma $ labels the twelve vectors connecting a Mo
site with any of its NN in the f.c.c. sublattice occupied by Mo atoms.

\begin{figure}
\narrowtext
\epsfxsize=3.5truein
\vbox{\hskip 0.05truein \epsffile{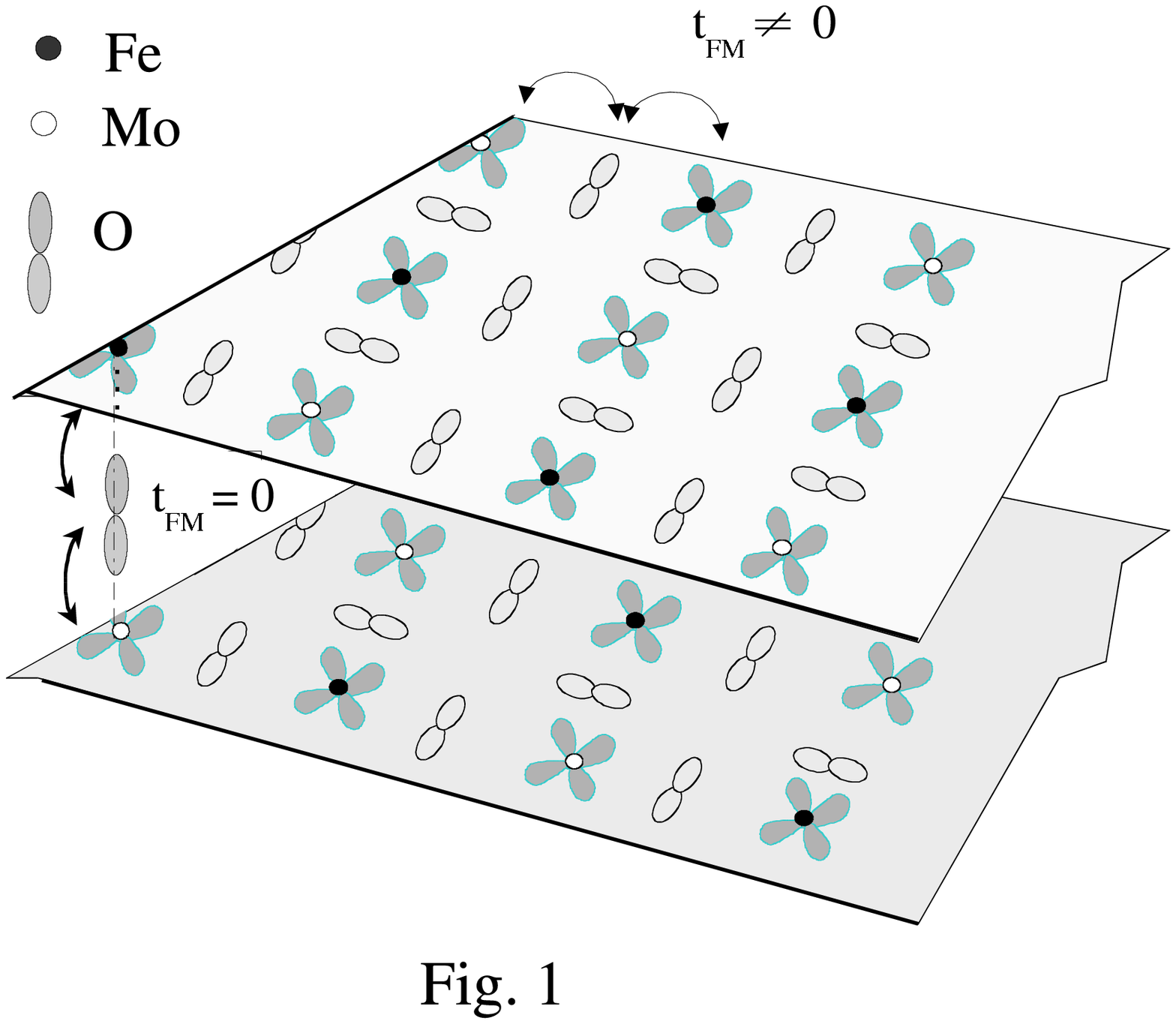}}
\medskip
\caption{Scheme of the effective hopping between nearest $d_{xy}$
orbitals through intermediate O $2p$ orbitals.}
\end{figure}

This is already an effective Hamiltonian in which the O atoms, which lie in
between any two transition metal atoms, were eliminated. This procedure can
be done using perturbation theory \cite{ali,Millis} if late 3d transition
metal atoms (like Ni or Cu) are not involved. Due to the symmetry of the
intermediate O orbitals, one realizes that hopping Fe-Mo $t_{FM}$ is two
dimensional: it is nonzero only between $\sigma $ orbitals lying on the
plane $\sigma $ (see Figure 1). Thus, it conserves color. We take $U_{Fe}=7$
eV from spectroscopic sudies.\cite{boc} Here we derive the other parameters
of the model for Sr$_{2}$FeMoO$_{6}$ by fitting {\it ab initio} results for
the $t_{2g}$ bands obtained previously,\cite{ali} to the corresponding
result of Eq. (\ref{hs}) in the Hartree-Fock approximation. In this
approximation $U_{Fe}$ is taken as zero and $\varepsilon _{Fe}$ is replaced
by $\varepsilon _{Fe}^{HF}=\varepsilon _{Fe}+\frac{2}{3}U_{Fe}n_{Fe}^{HF}$,
where $n_{Fe}^{HF}/3$ is the expectation value of $f_{i\sigma }^{\dagger
}f_{i\sigma }$ (independent of $i$ and $\sigma $) in the Hartree-Fock
approximation. We have adjusted the two eigenvalues for wave vector $\Gamma =%
{\bf 0}$, and the other two for wave vector $L=(\pi /a,\pi /a,\pi /a)$,
where $a$ is the lattice parameter of the f.c.c. structure. This fitting
gives the values of: $\varepsilon _{Fe}^{HF}$, $\varepsilon _{Mo}$, $t_{FM}$
and $t_{MM}(\sigma ,\sigma ,\gamma _{\sigma })$, where $\gamma _{\sigma }$
lies in the plane $\sigma $ (see Fig. 2). The other two independent values
of $t_{MM}(\sigma ,\sigma ^{\prime },\gamma )$ are smaller. For simplicity,
their values are derived using the geometrical relations that correspond to
direct Mo-Mo hybridization,\cite{slat} and taking the following relations
for the components with different angular momentum projection of the hopping
integrals: $(dd\delta )=0$, $(dd\pi )/(dd\sigma )=-0.54$.\cite{harr} The
occupation $n_{Fe}^{HF}$ was obtained from the resulting tight-binding
dispersion, what allows us to derive a $\varepsilon _{Fe}$ from $\varepsilon
_{Fe}^{HF}$. We have repeated the procedure for several lattice parameters.
Using the numerical derivative of the {\it ab initio} energy with respect to
volume, we obtain the dependence of the parameters with pressure, as
indicated in Table I. The most significant change with applied pressure is
the increase in the magnitude of $t_{FM}$. In contrast, $\varepsilon
_{Mo}-\varepsilon _{Fe}$ decreases with applied pressure. Both effects favor
a metallic state by decreasing the probability of finding localized
electrons at the Fe sites. Note that while $\varepsilon _{Fe}$ lies $\sim
2.5 $ eV below $\varepsilon _{Mo}$, $\varepsilon _{Fe}^{HF}$ is close to $%
\varepsilon _{Mo}$. This is consistent with other {\it ab initio}
calculations.\cite{koba,fang,sarma} As a consequence, $n_{Fe}^{HF}\sim 0.5$.
However this value is significantly increased, and the amount of covalency
is reduced, when $U_{Fe}$ is treated in a more realistic approximation.

\begin{figure}
\narrowtext
\epsfxsize=3.5truein
\vbox{\hskip 0.05truein \epsffile{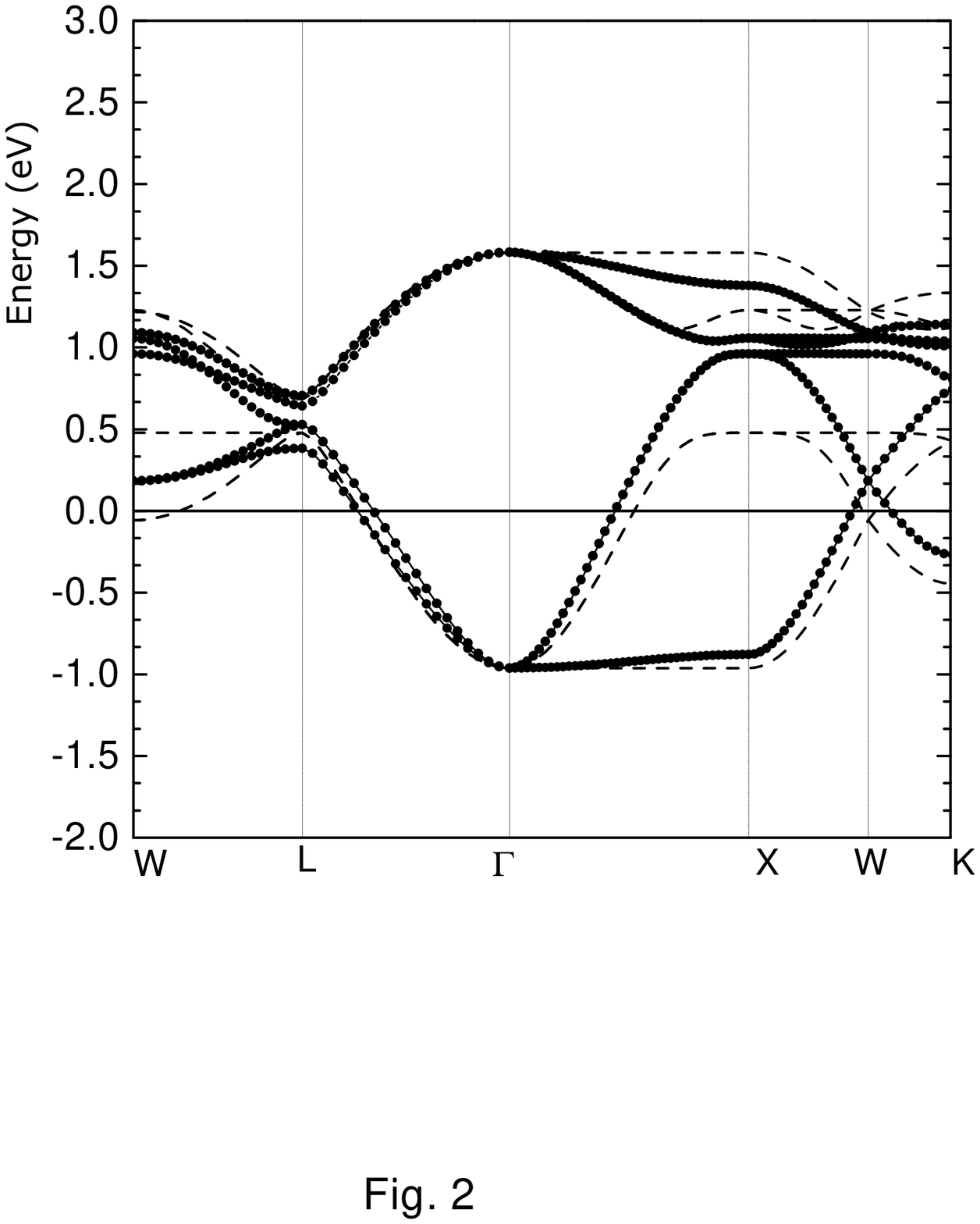}}
\medskip
\caption{Calculated $t_{2g}$ energy bands (full line and solid circles)
and fitting using Eq. (\ref{hs}) in the Hartree-Fock approximation (dashed
lines). The Fermi energy is at 0 eV. The wave vectors shown are: W$=(\pi
/a,0,2\pi /a)$, L$=(\pi /a,\pi /a,\pi /a)$, $\Gamma =(0,0,0)$, X$=(0,0,2\pi
/a)$, and K$=(3\pi /2a,3\pi /2a,0)$.}
\end{figure}

\begin{tabular}{|c|c|c|c|c|}
\hline
$a($\AA $)$ & $P($GPa$)$ & $t_{FM}$(eV) & $t_{MM}$(eV) & $\varepsilon
_{Mo}-\varepsilon _{Fe}$(eV) \\ \hline
{\ 16.51} & {\ -25.13} & {\ 0.1517} & {\ 0.0799} & {\ 2.787} \\ \hline
{\ 15.71} & {\ -24.95} & {\ 0.1955} & {\ 0.0956} & {\ 2.713} \\ \hline
{\ 15.31} & {\ -16.74} & {\ 0.2459} & {\ 0.1122} & {\ 2.623} \\ \hline
{\ 15.11} & {\ -9.14} & {\ 0.2781} & {\ 0.1222} & {\ 2.555} \\ \hline
{\ 14.91} & {\ -3.40} & {\ 0.2960} & {\ 0.1296} & {\ 2.512} \\ \hline
{\ 14.81} & {\ 3.45} & {\ 0.3150} & {\ 0.1361} & {\ 2.442 } \\ \hline
{\ 14.71} & {\ 7.04} & {\ 0.3242} & {\ 0.1390} & {\ 2.390} \\ \hline
{\ 14.61} & {\ 10.72} & {\ 0.3338} & {\ 0.1433} & {\ 2.268} \\ \hline
{\ 14.51} & {\ 18.15} & {\ 0.3510} & {\ 0.1531} & {\ 2.338} \\ \hline
\end{tabular}

Table I - Pressure and parameters of Eq. (\ref{hs}) for different lattice
parameters, obtained fiting {\it ab initio } energies at $\Gamma $ and L,
with the results of Eq. (\ref{hs}) in the Hartree-Fock approximation.

\section{The effective Hamiltonian}

In this section, we derive a Hamiltonian containing only effective Fe sites,
which describes the low-energy physics of Sr$_{2}$FeMoO$_{6}$ or other
double perovskites. The procedure is basically the same as the one used in
the cuprates \cite{jef,sch,sb,bel,tri,fei,ros,opt} and consists of the
following steps: i) change of basis of Mo $m_{i}$ orbitals to Wannier
functions centered at the Fe sites, ii) exact diagonalization of the cell
Hamiltonian $H_{i}$ at each Fe site $i$, retaining the ground state for each
number of particles. These states are mapped into the corresponding states
of a Hamiltonian that contains only Fe $t_{2g}$ states. iii) calculating the
intercell hopping matrix elements $(H-\sum_{i}H_{i})$ in the restricted
subspace of the cell ground states, and iv) inclusion of the excited states
by perturbation theory.\cite{tri,fei} We actually neglect the last step
because these corrections are very small, ensuring the validity of the
effective Hamiltonian.

\subsection{Change of basis of Mo orbitals}

The Mo Wannier orbitals $\alpha _{i\sigma }$ centered at Fe site $i$ are
constructed Fourier-transforming the $m_{j\sigma }$ in the plane $\sigma $
at which the orbital $\sigma $ lies, and then transforming the dependence on
the two-dimensional wave vector back to real space:

\begin{eqnarray}
\alpha _{i\sigma } &=&\frac{1}{N_{\sigma }}%
\sum_{k,l}e^{ik.(R_{i}-R_{l})}m_{l,\sigma }  \nonumber \\
&=&\sum_{\Delta _{\sigma }}C_{\sigma }\left( \Delta _{\sigma }\right) m_{i+{%
\Delta _{\sigma },\sigma }},  \nonumber \\
\text{with }C_{\sigma }\left( \Delta _{\sigma }\right)  &=&\frac{1}{%
N_{\sigma }}\sum_{k_{\sigma }}e^{-ik{_{\sigma }\Delta _{\sigma }}}
\label{alpha}
\end{eqnarray}
where $N_{\sigma }$ is the number of atoms lying in the plane $\sigma $, $%
k_{\sigma }$ are two-dimensional reciprocal lattice vectors parallel to the
plane $\sigma $, and $\Delta _{\sigma }$ are the vectors lying in the plane $%
\sigma $ which connect an Fe site with all Mo sites in this plane. The
vector $\Delta _{\sigma }$ does not belong to the Bravais lattice.
Performing the two-dimensional integral, one obtains:

\begin{equation}
C(\Delta _{\sigma })=\frac{4}{\pi ^{2}}\frac{(-1)^{n_{x}}-(-1)^{n_{y}}}{%
(n_{y}^{2})-(n_{x}^{2})}\text{, with }\Delta _{\sigma }=\frac{a}{2}(n_{x}%
{\bf \hat{x}}+n_{y}{\bf \hat{y}}),  \label{c}
\end{equation}
where $n_{x}$, $n_{y}$ are two integers of opposite parity and ${\bf \hat{x}}
$, ${\bf \hat{y}}$ are the unit vectors of the plane $\sigma $.

Using the inverse of Eq. (\ref{alpha}) and some algebra, the Hamiltonian in
the new basis takes the form:

\begin{eqnarray}
H &=&\sum_{i}H_{i}-(\sum_{i\sigma \mu _{\sigma }\neq 0}\widetilde{t_{FM}(}%
\sigma ,\mu _{\sigma })f_{i\sigma }^{\dagger }\alpha _{i+\mu _{\sigma
}\sigma }  \nonumber \\
&&+\sum_{i\sigma \sigma ^{\prime }\nu }\widetilde{t_{MM}}\left( \sigma
,\sigma ^{\prime },\nu \right) \alpha _{i\sigma }^{\dagger }\alpha _{i+\nu
\sigma ^{\prime }}+\text{H.c.}),  \label{ha}
\end{eqnarray}
with

\begin{eqnarray}
H_{i} &=&\varepsilon _{Fe}\sum_{\sigma }f_{i\sigma }^{\dagger }f_{i\sigma
}+\varepsilon _{Mo}\sum_{\sigma }\alpha _{i\sigma }^{\dagger }\alpha
_{i\sigma }  \nonumber \\
&&+U_{Fe}\sum_{\sigma <\sigma ^{\prime }}f_{i,\sigma }^{\dagger }f_{i,\sigma
}f_{i,\sigma ^{\prime }}^{\dagger }f_{i,\sigma ^{\prime }}+  \nonumber \\
&&-\sum_{\sigma }[\widetilde{t_{FM}}\left( \sigma ,0\right) f_{i\sigma
}^{\dagger }\alpha _{i\sigma }+\text{H.c.}].  \label{hi}
\end{eqnarray}
$\mu _{\sigma }$ and $\nu $ are vectors of the f.c.c. Bravais lattice and
the former lies in the plane $\sigma $. The change of basis has introduced
hoppings at arbitrary distances, but as we shall show, they decay very
rapidly with distance. This decay is different from those of similar hopping
terms in the case of the cuprates due to the different nature of the Wannier
orbitals. The hopping between Fe and new Mo orbitals at a distance $\mu
_{\sigma }=\frac{a}{2}(n_{x}{\bf \hat{x}}+n_{y}{\bf \hat{y}})$ (where now $%
n_{x}$, $n_{y}$ are two integers of the same parity) is:

\begin{eqnarray}
\frac{\widetilde{t_{FM}(}\sigma ,\mu _{\sigma })}{t_{FM}} &=&\sum_{\sigma
\delta _{\sigma }}C_{\sigma }(\mu _{\sigma }-\delta _{\sigma })  \nonumber \\
&=&\frac{16}{\pi ^{2}}\frac{(-1)^{n_{x}}}{%
(n_{x}^{2}-n_{y}^{2})^{2}-2(n_{x}^{2}+n_{y}^{2})+1}.  \label{tfm}
\end{eqnarray}
The weakest decay is along the diagonal direction $n_{x}=n_{y}$. For $%
n_{x}=n_{y}=0$, $\widetilde{t_{FM}}(\sigma ,0)/t_{FM}=16/\pi ^{2}\simeq 1.62$%
. As a consequence most of the original hopping terms are included in $%
\sum_{i}H_{i}$, that will be solved exactly. For the hopping between
different Mo $\alpha _{i}$ orbitals we have to distinguish two cases:

\begin{equation}
\widetilde{t_{MM}}{\left( \sigma ,\sigma ^{\prime },\nu \right) }=\left\{ 
\begin{array}{lr}
\ t_{MM}{(\sigma ,\sigma ,\nu )}\  & \sigma =\sigma ^{\prime } \\ 
\sum_{\eta }t_{MM}({\sigma ,\sigma ^{\prime },\nu -\eta )~}S{(}\eta ) & 
\sigma \neq \sigma ^{\prime }
\end{array}
\right. ,  \label{tmm}
\end{equation}
where

\begin{equation}
S(\eta )=\sum_{\Delta _{\sigma }}C_{\sigma }(\Delta _{\sigma })C_{\sigma
^{\prime }}(\Delta _{\sigma }+\eta ),  \label{s}
\end{equation}
and the sum over $\Delta _{\sigma }$ is restricted to vectors such that $%
\Delta _{\sigma }+\eta $ lies in the $\sigma ^{\prime }$ plane. This sum is
evaluated in Appendix A. The sum over $\eta $ in Eq. (\ref{tmm}) is
restricted to Bravais lattice vectors such that $\nu -\eta $ are NN in the
f.c.c. lattice, because otherwise $t_{MM}(\sigma ,\sigma ^{\prime },\nu
-\eta )$ vanishes. Also, due to symmetry arguments, if $\sigma \neq \sigma
^{\prime }$ then $t_{MM}(\sigma ,\sigma ^{\prime },\nu )=0$ unless $\nu $
lies in the plane perpendicular to both $\sigma $ and $\sigma ^{\prime }$.
The color-conserving hopping $\widetilde{t_{MM}}{\left( \sigma ,\sigma ,\nu
\right) }$ retains the same form as in the original Mo basis with a rigid
shift in the positions of the orbitals (from Mo to Fe positions). Some
values of these hoppings for the experimental lattice parameter are: $%
\widetilde{t_{FM}(}\sigma ,0)=$0.51 eV, $\widetilde{t_{FM}(}\sigma ,\delta
_{\sigma })=$0.17 eV, $\widetilde{t_{MM}}({\sigma ,\sigma ,\gamma _{\sigma
})=}$0.14 eV where ${\gamma _{\sigma }}$ connects nearest neighbor ions in
the f.c.c. lattice and lies in the plane $\sigma $. For larger distances, $%
\widetilde{t_{MM}}$ is at most of order 0.01 eV.

\subsection{Diagonalization of the cell Hamiltonian}

$H_{i}$ can be diagonalized exactly in each subspace of definite number of
electrons and of states which transform under the point group operations as
a basis function of some irreducible representation.. The ground state of
the most relevant subspaces is mapped into the corresponding state of a
monatomic f.c.c. lattice with $t_{2g}$ orbitals. For example, in the
subspace of one electron and wave functions transforming like $\sigma $ ($xy$%
, $yz$, or $zx$), the ground state has the form:

\begin{equation}
\left| i1\sigma \right\rangle =(A_{1}f_{i\sigma }^{\dagger }+B_{1}\alpha
_{i\sigma }^{\dagger })\left| 0\right\rangle ,  \label{1p}
\end{equation}
and is mapped onto the state $c_{i\sigma }^{\dagger }\left| 0\right\rangle $
of the effective monatomic Hamiltonian $H_{eff}$. Similarly, the ground
state for two particles has the form

\begin{eqnarray}
\left| i2\sigma \sigma ^{\prime }\right\rangle  &=&[A_{2}\alpha _{i\sigma
}^{\dagger }\alpha _{i\sigma ^{\prime }}^{\dagger }+B_{2}f_{i\sigma
}^{\dagger }f_{i\sigma ^{\prime }}^{\dagger }  \nonumber \\
&&+C_{2}(\alpha _{i\sigma }^{\dagger }f_{i\sigma ^{\prime }}^{\dagger
}+f_{i\sigma }^{\dagger }\alpha _{i\sigma ^{\prime }}^{\dagger })]\left|
0\right\rangle ,  \label{2p}
\end{eqnarray}
and is mapped onto the state $c_{i\sigma }^{\dagger }c_{i\sigma ^{\prime
}}^{\dagger }\left| 0\right\rangle $. In the present work and since we are
interested in occupations near one electron per lattice site, we disregard
the states with three electrons in the cell. This does not modify the
results for the metal-insulator transition presented in the next section.
The inclusion of three-particle states is straightforward but complicates
the notation and introduces several new terms in $H_{eff}$. For Sr$_{2}$FeMoO%
$_{6}$ at the experimentally observed lattice constant we obtain $%
E_{1}=\varepsilon _{Fe}-0.092eV$ and $E_{2}=2\varepsilon _{Fe}+2.442eV$,
where $E_{n}$ is the on-site energy in $H_{eff}$, and the effective Coulomb
repulsion is $U=E_{2}+E_{0}-2E_{1}=2.625eV$. This strong reduction with
respect to $U_{Fe}=7eV$ is due to the fact that $\left| i2\sigma \sigma
^{\prime }\right\rangle $ is dominated by the last term in Eq. (\ref{2p}).
In other words, it is energetically more favorable to occupy one Fe orbital
and one Wannier Mo orbital at that site, due to the large intratomic
repulsion in Fe.

\subsection{The intersite terms}

The hopping terms $H-\sum_{i}H_{i}$, calculated in the restricted basis
defined above, lead to effective hopping terms in $H_{eff}$. In addition to
the dependence on the lattice vector and orbitals involved, the effective
hopping depends on the occupation of the two sites involved. We denote by $%
t_{i}$ ($i=1$ to 3), the matrix elements which correspond to the following
transitions: 1) $\left| i0,j1\sigma \right\rangle \longleftrightarrow \left|
i1\sigma ^{\prime },j0\right\rangle $ 2) $\left| i1\sigma ^{\prime
},j1\sigma \right\rangle \longleftrightarrow \left| i0,j2\sigma \sigma
^{\prime }\right\rangle $ 3) $\left| i1\sigma _{1},j2\sigma _{2}\sigma
_{3}\right\rangle \longleftrightarrow \left| i2\sigma _{1}\sigma
_{4},j1\sigma _{2}\right\rangle $. The corresponding matrix elements turn
out to be: 
\begin{eqnarray*}
t_{1}(\sigma ,\sigma ^{\prime },\nu ) &=&2A_{1}A_{2}\widetilde{t_{FM}}%
(\sigma ,\sigma ^{\prime },\nu )+A_{1}^{2}\widetilde{t_{MM}}(\sigma ,\sigma
^{\prime },\nu ), \\
t_{2}(\sigma ,\sigma ^{\prime },\nu )
&=&[C_{2}(A_{1}^{2}+B_{1}^{2})+A_{1}B_{1}(A_{2}+B_{2})]\widetilde{t_{FM}}%
(\sigma ,\sigma ^{\prime },\nu ) \\
&&+A_{1}(A_{1}A_{2}+B_{1}C_{2})\widetilde{t_{MM}}(\sigma ,\sigma ^{\prime
},\nu ), \\
t_{3}(\sigma ,\sigma ^{\prime },\nu )
&=&2(A_{1}C_{2}+B_{1}B_{2})(A_{1}A_{2}+B_{1}C_{2})\widetilde{t_{FM}}(\sigma
,\sigma ^{\prime },\nu ) \\
&&+(A_{1}C_{1}+B_{1}C_{2})^{2}\widetilde{t_{MM}}(\sigma ,\sigma ^{\prime
},\nu ),
\end{eqnarray*}
where $\widetilde{t_{FM}}(\sigma ,\sigma ^{\prime },\nu )=\delta _{\sigma
\sigma ^{\prime }}\widetilde{t_{FM}}(\sigma ,\nu )$. The procedure followed
so far leads to the following effective Hamiltonian:

\begin{eqnarray}
&&H_{eff}=E_{1}\sum_{i\sigma }c_{i\sigma }^{\dagger }c_{i\sigma
}+U\sum_{i\sigma <\sigma ^{\prime }}c_{i,\sigma }^{\dagger }c_{i,\sigma
}c_{i,\sigma ^{\prime }}^{\dagger }c_{i,\sigma ^{\prime }}  \nonumber \\
&&-[\sum_{i\sigma \sigma ^{\prime }\nu }c_{i,\sigma }^{\dagger }c_{i+\nu
,\sigma ^{\prime }}\{t_{1}(\sigma ,\sigma ^{\prime },\nu )P_{i0}P_{i+\nu 1} 
\nonumber \\
&&+t_{2}(\sigma ,\sigma ^{\prime },\nu )(P_{i1}P_{i+\nu 1}+P_{i0}P_{i+\nu 2})
\nonumber \\
&&+t_{3}(\sigma ,\sigma ^{\prime },\nu )P_{i1}P_{i+\nu 2}\}+\text{H.c}.],
\label{heff}
\end{eqnarray}
where $P_{il}$ is the projector over $l$ number of particles at the
effective site $i$. Due to the rapid decay of the hopping terms with
distance, in what follows we shall retain only NN effective hopping. At this
distance, three hopping amplitudes $t_{j}$ can be distinguished for each sum
of occupations $j$: a) hopping which conserves color $(\sigma =\sigma
^{\prime })$ and the nearest neighbor vector $\gamma $ lies in the plane $%
\sigma $. We call this amplitude $t_{jc\parallel }$, b) same as before with $%
\gamma $ at an angle of $\pi /4$ with the plane $\sigma $. We call it $%
t_{jc\perp }$, and c) hopping that does not conserve color ($t_{jnc}$). In
this case for a hopping $\sigma \neq \sigma ^{\prime }$, $\gamma $ must lie
in the plane perpendicular to both $\sigma $ and $\sigma ^{\prime }$. The
largest of these nine hoppings are the $t_{jc\parallel }$. These are the
only ones which survive if the electronic stucture is approximated as
two-dimensional for each $\sigma $. For Sr$_{2}$FeMoO$_{6}$ we obtain at the
observed lattice parameter: $t_{1c\parallel }=0.197$ eV, $t_{2c\parallel
}=0.179$ eV, and $t_{3c\parallel }=0.109$ eV. The next two in decreasing
order of magnitude are $t_{1nc}=0.075$ eV and $t_{1c\perp }=0.047$ eV. The
remaining four hoppings lie below 0.03 eV.

\subsection{Corrections from excited states}

$H_{eff}$ can be systematically improved, including the effects of the
states neglected in Subsection {\bf B} by perturbation theory. The first
correction terms are of second order in effective hopping terms. Most of
them involve matrix elements of order 0.1 eV or smaller and denominators of
order 3eV or larger and can be neglected. The smallest denominator occurs
for an intermediate state created by a term similar to $t_{3}$, in which the
two-particle eigenstate of $H_{i}$ lies $\simeq 0.3eV$ above the ground
state. However for a non-interacting system, with one electron per site on
average, the probability of finding a singly occupied site is $\simeq 0.1$
and decreases with increasing $U$. Then, the effect of this correction on
the energy per site is less than $0.01eV$ and we can also neglect it for
fillings around one electron per site.

\section{The slave-boson mean-field treatment}

The formalism used here is essentially a generalization to more than two
colors (spin up and down) of the original formulation \cite{kot}. The
Hubbard model with orbital degeneracy was studied by Hasegawa \cite{hase}
extending a previous formalism used in the periodic Anderson model \cite{dor}%
, and by Fr\'{e}sard and Kotliar \cite{fres}. In our case, the interactions
include correlated hopping which does not conserve color. However, the
projectors $P_{il}$ are easily expressed in terms of bosonic operators, and
the approximation remains suitable for our problem. The basic idea is to
enlarge the Fock space to include bosonic states which correspond to each
state in the fermionic description. For example, the vacuum state at site $i$
is now represented as $e_{i}^{\dagger }\left| 0\right\rangle $, where $%
e_{i}^{\dagger }$ is a bosonic operator corresponding to the empty site;
similarly $s_{i\sigma }^{\dagger }c_{i\sigma }^{\dagger }\left|
0\right\rangle $ represents the simply occupied state with spin $\sigma $,
and so on. The bosonic operators for doubly and triply occupied states are
denoted $d_{i\sigma \sigma ^{\prime }}^{\dagger }$ and $t_{i}^{\dagger }$
respectively. In this way the projectors can be expressed in terms of boson
operators (for example $P_{i2}=\sum_{\sigma <\sigma ^{\prime }}d_{i\sigma
\sigma ^{\prime }}^{\dagger }d_{i\sigma \sigma ^{\prime }}$) and the
interactions between fermions disappear from the Hamiltonian. To restrict
the bosonic states to those with a physical meaning, the following
constraints must be satisfied:

\begin{eqnarray}
e_{i}^{\dagger }e_{i}+\sum_{\sigma }s_{i\sigma }^{\dagger }s_{i\sigma
}+\sum_{\sigma <\sigma ^{\prime }}d_{i\sigma \sigma ^{\prime }}^{\dagger
}d_{i\sigma \sigma ^{\prime }}+t_{i}^{\dagger }t_{i} &=&1  \nonumber \\
s_{i\sigma }^{\dagger }s_{i\sigma }+\sum_{\sigma ^{\prime }\neq \sigma
}d_{i\sigma \sigma ^{\prime }}^{\dagger }d_{i\sigma \sigma ^{\prime
}}+t_{i}^{\dagger }t_{i} &=&c_{i\sigma }^{\dagger }c_{i\sigma },
\label{cons}
\end{eqnarray}
To simplify the notation, we introduce the following operators: 
\begin{eqnarray}
X_{i\sigma }^{\dagger } &=&R_{i\sigma }s_{i\sigma }^{\dagger
}e_{i}L_{i\sigma },  \nonumber \\
Y_{i\sigma \sigma ^{\prime }}^{\dagger } &=&R_{i\sigma }d_{i\sigma \sigma
^{\prime }}^{\dagger }s_{i\sigma ^{\prime }}L_{i\sigma },  \label{xy}
\end{eqnarray}
where

\begin{eqnarray}
R_{i,\sigma } &=&(1-e_{i}^{\dagger }e_{i}-\sum_{\sigma ^{\prime }\neq \sigma
}s_{i,\sigma ^{\prime }}^{\dagger }s_{i,\sigma ^{\prime }}-d_{i,\sigma
^{\prime }\sigma ^{\prime \prime }}^{\dagger }d_{i,\sigma ^{\prime }\sigma
^{\prime \prime }})^{-1/2}\ ,  \nonumber \\
L_{i,\sigma } &=&(1-s_{i,\sigma }^{\dagger }s_{i,\sigma }-\sum_{\sigma
^{\prime }\neq \sigma }d_{i,\sigma \sigma ^{\prime }}^{\dagger }d_{i,\sigma
\sigma ^{\prime }})^{-1/2}.  \label{r}
\end{eqnarray}
with $\sigma ^{\prime }\neq \sigma \neq \sigma ^{\prime \prime }$. Inside
the expression (\ref{xy}) and using the first constraint (\ref{cons}) these
operators are strictly equal to 1. They are introduced to reproduce the
correct non-interacting result when the saddle-point approximation is made.%
\cite{ali,hase,dor,fres} The Hamiltonian takes the form

\begin{eqnarray}
\widehat{H}_{eff}^{SB} &=&E_{1}\sum_{i\sigma }c_{i\sigma }^{\dagger
}c_{i\sigma }+U\sum_{i\sigma \sigma ^{\prime }}d_{i,\sigma \sigma ^{\prime
}}^{\dagger }d_{i,\sigma \sigma ^{\prime }}  \nonumber \\
&&+\sum_{\left\langle ij\right\rangle \sigma ,\sigma ^{\prime }}\{c_{i\sigma
}^{\dagger }c_{j\sigma ^{\prime }}[t_{1}(\sigma ,\sigma ^{\prime
},R_{ij})X_{i\sigma }^{\dagger }X_{j\sigma ^{\prime }}+ \\
&&\sum_{\sigma _{1}}t_{2}(\sigma ,\sigma ^{\prime },R_{ij})(X_{i\sigma
}^{\dagger }Y_{j\sigma ^{\prime }\sigma _{1}}+Y_{i\sigma \sigma
_{1}}^{\dagger }X_{j\sigma ^{\prime }})  \nonumber \\
&&+\sum_{\sigma _{1}\sigma _{2}}t_{3}(\sigma ,\sigma ^{\prime
},R_{ij})Y_{i\sigma \sigma _{1}}^{\dagger }Y_{j\sigma ^{\prime }\sigma
_{2}}]+\text{H.c}.\}  \label{hsb}
\end{eqnarray}
In the saddle-point approximation for the uniform, color-independent
solution, all bosonic operators are condensed, {\it i.e.} replaced by
numbers, independent of position and color ($e_{i}^{\dagger }=e$, $%
s_{i\sigma }^{\dagger }=s$, $d_{i\sigma \sigma ^{\prime }}^{\dagger }=d$, $%
t_{i}^{\dagger }=t$), and their values are obtained minimizing the energy of
the resulting non-interacting fermionic problem under the constraints (\ref
{cons}). For one electron per site in the insulating state, one has $s=1/3$, 
$e=d=t=0$. In general, for a multicolor problem, the condensates for only
one occupation $n$ are different from zero, and near the metal-insulator
transition, the values for $n-1$ and $n+1$ are infinitesimals of the same
order, while the other expectation values of the condensates are
infinitesimals of larger order. Thus, using the constraints we can write
near the metal-insulator transition:

\begin{equation}
e^{2}=3d^{2}\text{; }s^{2}=\frac{1}{3}-2d^{2}\text{; }t=0,  \label{bt}
\end{equation}
where $d\rightarrow 0$. The stability with respect to $d$ determines the
position of the metal-insulator transition. Specifically, replacing (\ref{bt}%
) in (\ref{hsb}) the energy up to order $d^{2}$ takes the form 
\begin{equation}
E(d)=E_{1}+(3U+E_{TB})d^{2}  \label{ed}
\end{equation}
where $E_{TB}$ is the energy of a tight-binding Hamiltonian in the f.c.c.
lattice, in which the three different NN hopping amplitudes $T_{\xi }$
(denoted by the subscripts $\xi =c\parallel $, $c\perp $ or $nc$ are
described as before) are weighted averages of the $t_{i\xi }$:

\begin{equation}
T_{\xi }=t_{1\xi }+\frac{4\sqrt{3}}{3}t_{2\xi }+\frac{4}{3}t_{3\xi }
\label{ttb}
\end{equation}

Clearly, the transition is at the point $U=-E_{TB}/3$, and for the case in
which the hoppings do not depend on the occupancy of the sites involved,
previous results are recovered.\cite{hase,fres} We have calculated $E_{TB}$
using a mesh of 286 points in $1/48$ of the Brillouin zone of the f.c.c.
lattice. The values of $t_{i\xi }$ and $U$ were calculated as described in
the previous section, and we have looked for the value of $\varepsilon
_{Mo}-\varepsilon _{Fe}$ (the most uncertain parameter of the original
Hamiltonian) at which the metal-insulator transition takes place. The result
as a function of pressure is represented in Fig. 3. In the same figure we
show the value of $\varepsilon _{Mo}-\varepsilon _{Fe}$ for Sr$_{2}$FeMoO$%
_{6}$ derived as explained in Section II from {\it ab initio} calculations.
If for a given pressure this {\it ab initio} value lies below the
metal-insulator boundary obtained with the SBMFA, the system is predicted to
be a metal. In the opposite case, the mobility of the carriers is strongly
reduced as a consequence of the electronic correlations and an insulating
behavior is expected. We actually see that Sr$_{2}$FeMoO$_{6}$ is very near
a metal-insulator transition and that negative pressure can drive it
insulating, mainly due to reduction of the hopping amplitudes as the lattice
parameter is expanded. This might be the main reason of the insulating
character of Sr$_{2}$FeWO$_{6}$. In our formalism, in the insulating phase,
the occupation at Fe is given by $A_{1}^{2}$. This value at the transition
is $0.97$ and should decrease further when delocalization effects, not
adequately taken into account by the SBMFA, are included. Then, one expects
an oxidation state near $+2$ for Fe in the insulating state, but not exactly
Fe$^{+2}$. Similarly, the magnetic moment at Fe sites is slightly above $%
4\mu _{B}$, and that of Mo slightly below zero. Covalency with O atoms,
modifies these values, particularly the valence.\cite{ali}

\begin{figure}
\narrowtext
\epsfxsize=3.5truein
\vbox{\hskip 0.05truein \epsffile{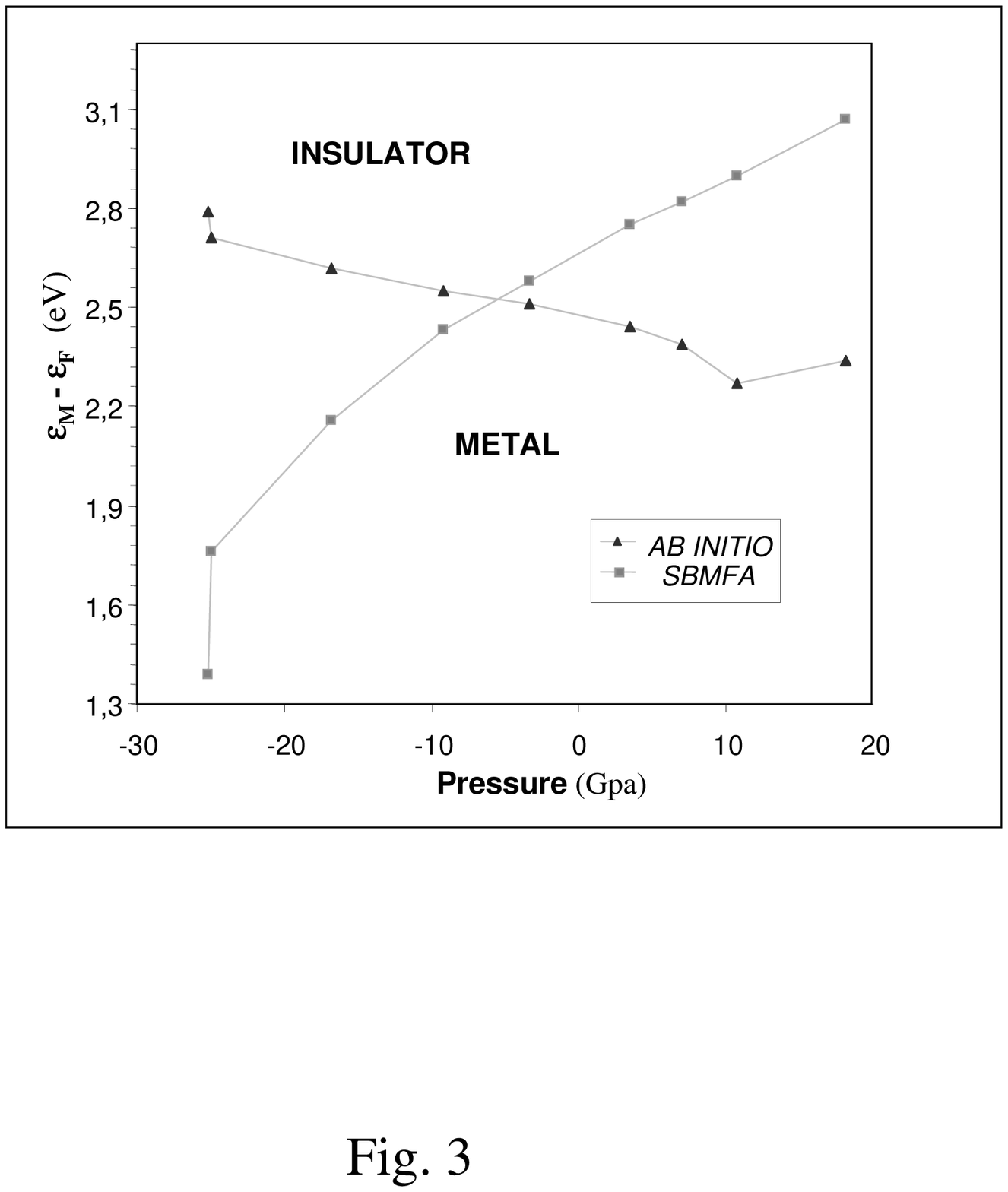}}
\medskip
\caption{Full line: phase diagram for the metal-insulator transition as
a function of $\varepsilon _{Mo}-\varepsilon _{Fe}$ and pressure. Dashed
line: values of $\varepsilon _{Mo}-\varepsilon _{Fe}$  for Sr$_{2}$FeMoO$_{6}
$, obtained using the {\it ab initio} results of Table I.}
\end{figure}

\section{Summary and discussion}

Starting from a model describing transition-metal $t_{2g}$ orbitals in
half-metallic double perovskites like Sr$_{2}$FeMoO$_{6}$, and assuming that
on-site Coulomb interactions are important in only one of the
transition-metal ions (Fe), we have derived an effective Hamiltonian $%
H_{eff} $, which contains only one species of effective atoms in an f.c.c.
lattice with moderate on-site interaction. This seems to be the maximum
possible reduction of the relevant Hilbert space per unit cell, and we
expect that it can be useful for future studies in similar systems in which
correlations are explicitly taken into account. As long as the difference in
bare on-site energies (in contrast to the Hartree-Fock or {\it ab initio}
ones) is not very small, the interactions in $H_{eff}$ are important and
some treatment which appropriately takes into account many-body effects is
necessary for its study. Our derivation can be generalized to include both
spins. However, our formalism is not appropriate to treat the case of
disorder systems in which the highly correlated transition metal can occupy
both sublattices.\cite{saha} For an occupation near one $t_{2g}$ minority
electron per unit cell, $H_{eff}$ is very accurate. If the occupation is
near two electrons per site, our calculations should be extended to include
states with three electrons per unit cell, and perturbative corrections due
to neglected excited states in the two particle sector.

We have applied $H_{eff}$ to analyze the metal-insulator transition as a
function of pressure or $\varepsilon _{Mo}-\varepsilon _{Fe}$ in Sr$_{2}$%
FeMoO$_{6}$. While a similar study was previously done,\cite{ali} the
accuracy of the present results should be higher because the highest energy
involved in the problem ($U_{Fe}\simeq 7eV$) and most of the Fe-Mo hopping,
is treated exactly in $H_{eff}$. We obtain that the region of insulating
behavior increases with respect to the previous calculation, and a
metal-insulator transition is predicted for a negative pressure near -5
GPa.. Although we do not know the parameters of the starting model for Sr$%
_{2}$FeWO$_{6}$, we believe that the insulating character of the compound,%
\cite{w} or the alloys Sr$_{2}$FeW$_{x}$Mo$_{1-x}$O$_{6}$ for $x\simeq 1$ is
related with this transition. $\varepsilon _{W}-\varepsilon _{Fe}$ is
expected to be higher than $\varepsilon _{Mo}-\varepsilon _{Fe}$ due to the
larger W-O hopping.\cite{fang} Actually for $x=1$, the system is
antiferromagnetic and our $H_{eff}$ derived here does include both spins.
However, we believe that the difference in energy between ferro- and
antiferromagnetic insulating phases is due to double exchange interactions
which are smaller than the energetic ingredients, whose competition
determines the metal-insulator transition. In simple terms, in the metallic
state, ferromagnetism is favored by energies of the order of the effective $%
t_{j}\sim 0.2$ eV (similarly to other proposals \cite{fang,kana}), while in
the insulating state, antiferromagnetism is favored by energies of order $%
t_{2}^{2}/U\sim 0.01$ eV. The situation is similar to the metal-insulator
transition which takes place in $R$NiO$_{3}$ replacing rare earths $R$ or
changing temperature.\cite{torr} It seems that the antiferromagnetic order
does not affect the metal-insulator transition, and the boundary between
metal and paramagnetic or antiferromagnetic order is continuous and smooth.
In the starting Hamiltonian we have assumed that the O atom lies in between
its two NN transition-metal (TM) ions. If this is not the case, as suggested
by some structural measurements \cite{mori} an increase in lattice parameter
might increase the effective hopping between TM atoms (for example by a
reduction of the TM-O-TM angle, as in RNiO$_{3}$ \cite{torr}). This might be
the reason Ba$_{2}$FeMoO$_{6}$ is still metallic in spite of the increase in
the distance between TM atoms.

\section*{Acknowledgments}

A. A. A. is partially supported by CONICET. This work was sponsored by PICT
03-06343 from ANPCyT and PIP 4952/96 of CONICET.

\appendix 

\section{Evaluation of the sums $S$}

Here we describe the evaluation of the sums $S(\eta )=\sum_{\Delta _{\sigma
}}C_{\sigma }(\Delta _{\sigma })C_{\sigma ^{\prime }}(\Delta _{\sigma
^{\prime }})$, with $\Delta _{\sigma ^{\prime }}-\Delta _{\sigma }=\eta $
and $\sigma \neq \sigma ^{\prime }$. For the sake of clarity, we assume $%
\sigma =xy$ and $\sigma ^{\prime }=zx$. The other cases are easily obtained
using symmetry arguments. We can then write:

\begin{eqnarray}
\eta &=&\frac{a}{2}(n_{x}{\bf \hat{x}}+n_{y}{\bf \hat{y}+}n_{z}{\bf \hat{z}})%
\text{; }\Delta _{xy}=\frac{a}{2}(n_{x}^{\prime }{\bf \hat{x}}-n_{y}{\bf 
\hat{y}});  \nonumber \\
\Delta _{zx} &=&\frac{a}{2}[(n_{x}^{\prime }+n_{x}){\bf \hat{x}}+n_{z}{\bf 
\hat{z}],}  \label{a1}
\end{eqnarray}
where $n_{x}${\bf , }$n_{x}^{\prime }$, $n_{y}$, and $n_{z}$ are integers
with $n_{x}+n_{y}{\bf +}n_{z}$ even and $n_{x}^{\prime }+n_{y}$ odd ($\eta $
is a vector of the f.c.c. lattice, but $\Delta _{\sigma }$ and $\Delta
_{\sigma ^{\prime }}$ are not). Using the definition of $\Delta _{\sigma }$
(Eq.(\ref{alpha})), we can write:

\begin{eqnarray}
S &=&\sum_{n_{x}^{\prime }}\frac{1}{N_{xy}N_{yz}}\sum_{k_{x},k_{y},k_{x}^{%
\prime },k_{z}^{\prime }}  \nonumber \\
&&\exp [i\frac{a}{2}(-k_{x}n_{x}^{\prime }+k_{y}n_{y}+k_{x}^{\prime
}(n_{x}^{\prime }+n_{x})+k_{z}^{\prime }n_{z}].  \label{sf}
\end{eqnarray}
We can assume that the system is a cube of edge $L$, and then the number of
atoms lying in the plane $\sigma $ is $N_{\sigma }=2(L/a)^{2}$. The sums
over the two dimensional wave vectors $k$ and $k^{\prime }$ run over a
square Brillouin zone with sides of length $2\sqrt{2}\pi /a$ rotated $\pi /4$
with respect to ${\bf \hat{x}}$. Depending on the parity of $n_{y}$, the sum
over $n_{x}^{\prime }$ runs either over all odd values or over all even
values, and:

\begin{equation}
\sum_{n_{x}^{\prime }}\exp [i\frac{a}{2}n_{x}^{\prime }(k_{x}^{\prime
}-k_{x})]=\frac{L}{a}[\delta _{k_{x},k_{x}^{\prime }}-(-1)^{n_{y}}\delta
_{k_{x},k_{x}^{\prime }+2\pi /a}].  \label{a3}
\end{equation}
Using this expression to eliminate the sum over $k_{x}$ in Eq. (\ref{sf}),
and replacing the remaining sums by integrals in reciprocal space, one
obtains:

\begin{eqnarray}
S(\eta ) &=&\frac{a^{3}}{4\pi ^{3}}\int_{0}^{2\pi /a}dk_{x}^{\prime }\cos
(k_{x}^{\prime }n_{x})\int_{0}^{2\pi /a-k_{x}^{\prime }}dk_{z}^{\prime }\cos
(k_{z}^{\prime }n_{x})  \nonumber \\
&&\times [\int_{0}^{2\pi /a-k_{x}^{\prime }}dk_{y}\cos (k_{y}n_{y}) 
\nonumber \\
&&-(-1)^{n_{y}}\int_{0}^{k_{x}^{\prime }}dk_{y}\cos (k_{y}n_{y})].
\label{integ}
\end{eqnarray}
The integrals are elementary and after some algebra, the result becomes:

\begin{eqnarray}
S\left( \frac{a}{2}(n_{x}{\bf \hat{x}}+n_{y}{\bf \hat{y}+}n_{z}{\bf \hat{z}}%
)\right) =\frac{(-1)^{n_{x}}}{\pi ^{2}n_{y}n_{z}}(\delta _{n_{x}+n_{y}
-n_{z},0}  \nonumber \\
+\delta _{n_{x}-n_{y}+n_{z},0}-\delta _{n_{x}+n_{y}+ n_{z},0}-\delta
_{-n_{x}+n_{y} +n_{z},0}).  \label{a5}
\end{eqnarray}

This expression and Eqs (\ref{tfm}) and (\ref{tmm}), allow to calculate all
hopping terms in the new basis.


\begin{references}
\bibitem{koba}  K.I. Kobayashi, T. Kimura, H. Sawada, K. Terakura and Y.
Tokura, Nature {\bf 395}, 677 (1998).

\bibitem{ba}  A. Maignanan, B. Raveau, C. Martin, and M. Hervieu, J. Solid
State Chem. {\bf 144}, 244 (1999).

\bibitem{re}  K. I. Kobayashi, T. Kimura, Y. Tomioka, H. Sawada, K.
Terakura, and Y. Tokura, Phys. Rev. B {\bf 59}, 11159 (1999).

\bibitem{w}  H. Kawanaka, I. Hase, S. Toyama, and Y. Nishihara, J. Phys.
Soc. Jpn. {\bf 68}, 2890 (1999).

\bibitem{alloy}  K.I. Kobayashi, T. Okuda, Y. Tomioka, T. Kimura, and Y.
Tokura, J. Magn. Magn. Mat. {\bf 218}, 17 (2000); R.I. Dass and J.B.
Goodenough, Phys. Rev. B {\bf 63}, 064417 (2001).

\bibitem{itoh}  M. Itoh, I. Ohta, and Y. Inaguma, Mat. Science Ing. B {\bf 41%
}, 55 (1996).

\bibitem{fang}  Z. Fang, K. Terakura, and J. Kanamori, Phys. Rev. B {\bf 63}%
, 180407 (2001).

\bibitem{ali}  A.A. Aligia., P. Petrone, J. Sofo, and B. Alascio, Phys. Rev.
B {\bf 64}, 092414 (2001).

\bibitem{more}  M.S. Moreno, J.E. Gayone, A. Caneiro, D. Niebieskiwiat, R.D.
Sanchez, and G. Zampieri, Solid State Commun. {\bf 120}, 161 (2001).

\bibitem{lind}  J. Lind\'{e}n, T. Yamamoto, M. Karppinen, and H. Yamauchi,
Appl. Phys. Lett. {\bf 76}, 2925 (2000).

\bibitem{gar}  B. Garc\'{\i }a-Landa, C. Ritter, M.R. Ibarra, J. Blasco,
P.A. Algarabel, R. Mahendiran, and J. Garc\'{i}a, Solid State Commun. {\bf %
110}, 435 (1999).

\bibitem{kot}  G. Kotliar and A.E. Ruckenstein, Phys. Rev. Lett. {\bf 57},
1362 (1986).

\bibitem{zr}  F.C. Zhang and T. M. Rice, Phys. Rev. B {\bf 37}, 3759 (1988).

\bibitem{jef}  J. H. Jefferson, H. Eskes, and L. F. Feiner, Phys. Rev. B 
{\bf 45}, 7959 (1992).

\bibitem{sch}  H.B. Schuttler and A.J. Fedro, Phys. Rev. B {\bf 45}, 7588
(1992).

\bibitem{sb}  M. E. Simon and A.A. Aligia, Phys. Rev. B {\bf 48}, 7471
(1993).

\bibitem{bel}  V.I. Belinicher and A. L. Chernyshev, Phys. Rev. B {\bf 49},
9746 (1994).

\bibitem{tri}  M. E. Simon and A.A. Aligia, Phys. Rev. B {\bf 52}, 7701
(1995).

\bibitem{fei}  L.F. Feiner, J.H. Jefferson and R. Raimondi, Phys. Rev. B 
{\bf 53}, 8751 (1996).

\bibitem{ros}  H. Rosner, H. Eschrig, R. Hayn, S.-L. Drechsler, and J.
M\'{a}lek, Phys. Rev. B{\it \ }{\bf 56}, 3402 (1997); references therein.

\bibitem{opt}  M. E. Simon, A.A. Aligia., and E.R. Gagliano, Phys. Rev. B 
{\bf 56}, 5637 (1997); references therein.

\bibitem{roze}  M.J. Rozenberg, X.Y. Zhang, and G. Kotliar, Phys. Rev. Lett. 
{\bf 69}, 1236 (1992).

\bibitem{geor}  A. Georges and W. Krauth, Phys. Rev. Lett. {\bf 69}, 1240
(1992).

\bibitem{Millis}  A. Chattopadhyay and A. J. Millis, Phys. Rev. B {\bf 64},
024424 (2001).

\bibitem{boc}  A.E. Bocquet, T. Mizokawa, T. Saitoh, H. Namatame, and A.
Fujimori, Phys. Rev. B {\bf 46}, 3771 (1992).

\bibitem{slat}  J. C. Slater and G. F. Koster, Phys. Rev. {\bf 94}, 1498
(1954); R.R. Sharma, Phys. Rev. B {\bf 19}, 2813 (1979).

\bibitem{harr}  W.A. Harrison, {\it Electronic structure and the Properties
of Solids} (W.H. Freeman and Co. San Francisco, 1980).

\bibitem{sarma}  D.D Sarma, P. Mahadevan, T. Saha-Dasgupta, S. Ray, and A.
Kumar, Phys. Rev. Lett. {\bf 85}, 2549 (2000).

\bibitem{hase}  H. Hasegawa, J. Phys. Soc. Jpn. {\bf 66}, 1391 (1997).

\bibitem{dor}  V. Dorin and P. Schlottmann, Phys. Rev. B {\bf 47}, 5095,
(1993).

\bibitem{fres}  R. Fr\'{e}sard and G. Kotliar, Phys. Rev. B {\bf 56}, 12909
(1997).

\bibitem{saha}  T. Saha - Dasgupta and D.D Sarma, Phys. Rev. B {\bf 64},
064408 (2001).

\bibitem{kana}  J. Kanamori and K. Terakura, , J. Phys. Soc. Jpn. {\bf 70},
1433 (2001).

\bibitem{torr}  J. B. Torrance, P. Lacorre, A. I. Nazzal, E.J. Ansaldo, and
Ch. Niedermayer, Phys. Rev. B 45, 8209 (1992).

\bibitem{mori}  Y. Moritomo, Sh. Xu, A. Machida, T. Akimoto, E. Nishibori,
M. Takata, and M. Sakata, Phys. Rev. B {\bf 61}, R7827 (2000).
\end{references}
\end{document}